\documentclass[runningheads]{llncs}
\usepackage{times}
\normalfont

\usepackage{centernot}
\usepackage{xcolor}
\usepackage{booktabs} 
\usepackage{paralist} 
\usepackage{xcolor,pifont}
\usepackage{colortbl}
\usepackage{multirow} 

\usepackage{amsfonts} 
\usepackage{amssymb} 
\usepackage{amsmath}
\usepackage{wasysym}

\usepackage{thm-restate}
\usepackage{etoolbox}
\newtoggle{APPENDIX}
\toggletrue{APPENDIX}  

\usepackage{amstext,latexsym}
\usepackage[T1]{fontenc}
\usepackage{eurosym}
\usepackage[pdftex]{graphicx}
\newcommand{\Bbox}{{\Box}}
\newcommand{\Opt}{{\scalebox{1.1}{$\circ$}}}
\usepackage{url}

\usepackage{algorithm,algpseudocode}

\usepackage{tikz}
\usetikzlibrary{arrows,calc,shapes.arrows,shapes.multipart,shapes.geometric, positioning, shadows, automata,fit,decorations.markings,decorations.pathmorphing,backgrounds,fadings}
\tikzset{elliptic state/.style={draw,ellipse}}

\makeatletter
\RequirePackage[bookmarks,unicode,colorlinks=true]{hyperref}%
   \def\@citecolor{blue}%
   \def\@urlcolor{blue}%
   \def\@linkcolor{blue}%

\def\orcidID#1{\smash{\href{http://orcid.org/#1}{\protect\raisebox{-1.25pt}{\protect\includegraphics{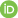}}}}}
\makeatother

\begin{document}

\title{Non-reducible Modal Transition Systems}
\author{Davide  Basile\orcidID{0000-0002-2930-6367} }
\authorrunning{Davide Basile}

\institute{ISTI--CNR, Pisa, Italy}

\maketitle

\begin{abstract}
Modal Transition Systems (MTS) are a well-known formalism that extend Labelled Transition Systems (LTS) with the possibility of specifying necessary and permitted behaviour. 
Modal refinement ($\preceq_m$) of MTS represents a step of the design process, namely the one in which some optional behaviour is discarded while other optional behaviour becomes necessary. 
Whenever two MTS are not in modal refinement relation, it could still be the case that the set of implementations of one MTS is included in the set of implementations of the other.   
 The challenge of devising an alternative notion of modal refinement that is both sound and complete with respect to the set of implementations, without disregarding valuable implementations, remains open. 
We introduce a subset of MTS called Non-reducible Modal Transition Systems (NMTS), together with a novel refinement relation $\preceq_n$ for NMTS. 
We illustrate through examples the additional constraints imposed by NMTS. 
Furthermore, we discuss a property holding for NMTS whose implementations are non-deterministic. 
\end{abstract}

\section{Introduction}

Modal Transition Systems (MTS)~\cite{LT88} extend Labelled Transition Systems (LTS)~\cite{Kel76} by distinguishing two
types of transitions, meant to describe necessary and optional behaviour in a system specification by means of transitions that must \emph{necessarily} be implemented and transitions that may \emph{optionally} be implemented. 
MTS come with a concept of \emph{refinement}, which represents a step of the design process, namely the one in which some optional behaviour is discarded while other optional behaviour becomes necessary. Stepwise refinement of an MTS eventually results in an \emph{implementation}, which is an LTS in which no further refinement is possible.
Refinement of MTS is critical for enabling formal reasoning on the correctness of a system's design and implementation, by enabling gradually refining an abstract specification into a concrete one, ensuring that each step is correct. 
MTS are a well-known specification theory and significant advances have been made so far~\cite{AHLNW08,Kre17}.
 
It is known that the (modal) refinement of MTS is not complete (cf., e.g.,~\cite{LNW07c}). 
In other words, there are cases in which two MTS are not in a refinement relation although the set of implementations of one MTS is included in the set of implementations of the other MTS (this relation is known as thorough refinement). 
Furthermore, while determining MTS refinement can be computed in polynomial time, determining thorough refinement of MTS requires EXPTIME~\cite{BKLS12}. 
%
%
In~\cite{LNW07c}, the problem of proposing an alternative notion of modal refinement that is both sound and complete with respect to its set of implementations is left open. It is also essential to demonstrate that the considered set of implementations is also interesting from a practical point of view (i.e., no valuable implementation is disregarded).

In this working paper\footnote{Please note that this paper is in draft version, so there could be possible typos and errors.}, 
we propose a subset of MTS, called \emph{Non-reducible Modal Transition Systems} (NMTS) together with their alternative notion of modal refinement $\preceq_n$. 
A fundamental insight behind NMTS is that states non-deterministically  reachable through the  execution of identical sequences of actions are related. Specifically, the outgoing transitions sharing the same action label also share the same modality. 
Furthermore, when a refinement step deactivates one optional transition, this leads to the deactivation of all other transitions that share the same label from all other related (source) states. 

The contributions of this paper are:
\begin{enumerate}
\item we introduce NMTS, a subset of MTS. In NMTS, the  transitions sharing the same action label are constrained  to also share the same modality whenever they are reachable by the same sequence of actions;
\item we equip NMTS with an alternative notion of modal refinement, called NMTS refinement. The refinement of NMTS is derived from modal refinement by imposing an additional constraint on the optional transitions of the system to be refined;
\item we provide different examples of MTS instances that fail to meet the requisites for being  either NMTS or refinements of NMTS; 
%
\item we introduce the  \emph{non-reducible non-determinism} property concerning optional, non-deterministic  actions. This non-determinism is inherent in such actions and should be preserved in any implementation where the action remains active.
All implementations accepted by the standard MTS refinement, but discarded by the NMTS refinement, are  showed to be implementations violating the non-reducible non-determinism property.
\item we provide an example (Figure~\ref{fig:notransitivity}) demonstrating that NMTS refinement is not transitive, and wherein two systems are in NMTS refinement despite their implementations not being in a relation of set inclusion.
\end{enumerate}

\paragraph{Overview}
Section~\ref{sect:background} introduces background on MTS and modal refinement. 
Section~\ref{sect:refinement} presents Non-reducible MTS (NMTS) and their refinement. Section~\ref{sect:example} discusses the property of non-reducible non-determinism, showing that the implementations discarded by NMTS refinement but accepted by modal refinement are violating this property. 
Section~\ref{sect:relatedwork} discusses the related work, while Section~\ref{sect:conclusion} concludes the paper and discusses future work.

\section{Background}\label{sect:background}

We start by discussing some background on MTS.
The standard definition of MTS accounts for two sets of transitions, \emph{permitted} (or \emph{may}) transitions, denoted by $\Delta_\Diamond$, and \emph{necessary} (or \emph{must}) transitions, denoted by $\Delta_{\Box}$, such that $\Delta_{\Box} \subseteq \Delta_\Diamond$, i.e., all (necessary) transitions are permitted. 
A transition $(q,a,q')\!\in\!\Delta_\Diamond$ is also denoted as $q\xrightarrow{a}_\Diamond q'$ and likewise $q\xrightarrow{a}_{\Box} q'$ if $(q,a,q')\!\in\!\Delta_{\Box}$. 
The reader may be misled to think that $q\xrightarrow{a}_\Diamond q'$ excludes $q\xrightarrow{a}_{\Box} q'$, 
and vice versa that $q\xrightarrow{a}_{\Box} q'$ excludes $q\xrightarrow{a}_\Diamond q'$. 
However, the first statement is not always true and the second is always false, since $\Delta_{\Box}\!\subseteq\!\Delta_\Diamond$. 
For our purpose, it is irrelevant to indicate that a transition is permitted. 
For the sake of simplifying the presentation, we thus opt for a slightly revised definition of MTS, where we partition the set of transitions into \emph{optional} and \emph{necessary} transitions, and no longer indicate the fact that all transitions are \emph{permitted}.

\begin{definition}[MTS]\label{def:MTS}
A \emph{Modal Transition System (MTS)} $S$ is a 5-tuple \linebreak 
$S=(Q, A, \overline{q}, \Delta_\Opt, \Delta_{\Box})$,
with set $Q$ of states, set $A$ of actions, initial state $\overline{s}\in Q$, and 
transition relation $\Delta \subseteq Q \times A \times Q$ partitioned into \emph{optional transitions}, 
denoted by~$\Delta_\Opt$, and \emph{necessary transitions}, denoted by~$\Delta_{\Box}$,
i.e., $\Delta_{\Opt} \cap \Delta_{\Box} = \emptyset$.
If $(s,a,s')\in\Delta_\Opt$, then we also write $s\xrightarrow{a}_\Opt s'$, and likewise we also write $s\xrightarrow{a}_{\Box} s'$ for $(s,a,s')\in\Delta_{\Box} $. 
We write  $s\xrightarrow{a} s'$  when $(s,a,s')\in\Delta$. 
We may omit the target state when it is immaterial. 
\end{definition}

Note that the standard definition of MTS is $(Q, A, \overline{q}, \Delta_\Diamond, \Delta_{\Box})$, where $\Delta_\Diamond=\Delta_\Opt \cup \Delta_{\Box}$. 
An LTS is an MTS where $\Delta_\Opt=\emptyset$.
In the sequel, the conversion from an MTS (and NMTS, cf. Section~\ref{sect:refinement}) $(Q, A, \overline{q}, \Delta_\Opt, \Delta_{\Box})$  with  $\Delta_\Opt = \emptyset$ to an LTS $(Q, A, \overline{q}, \Delta)$ with $\Delta = \Delta_{\Box}$ is implicit.
Moreover, we will use subscripts or superscripts to indicate the 
origin of an element of a tuple, i.e., 
$S = (Q_S, A_S, \overline{s}, \Delta^\Opt_S, \Delta^\Box_S)$. 
We now define modal refinement of MTS.

\begin{definition}[modal refinement]\label{def:modrefinement}
An MTS $S$ is a \emph{(modal) refinement\/} of an MTS $T$, denoted by $S\preceq_{m} T$, if and only if there exists a \emph{refinement relation\/} ${\cal R}\subseteq Q_S\times Q_T$ such that $(\overline{s},\overline{t})\in{\cal R}$ and for all $(s,t)\in{\cal R}$, the following holds:
\begin{enumerate}
\item whenever $t\xrightarrow{a}_{\Bbox} t'$, for some $t'\!\in\!Q_T$ and $a\!\in\!A_T$, then $a\!\in\!A_S$, $\exists\,s'\!\in\!Q_S: s \xrightarrow{a}_{\Bbox} s'$, and $(s',t')\in{\cal R}$, and
\item whenever $s\xrightarrow{a} s'$, for some $s'\!\in\!Q_S$ and $a\!\in\!A_S$, then $a\!\in\!A_T$, $\exists\,t'\!\in\!Q_T: t\xrightarrow{a} t'$, and $(s',t')\in{\cal R}$.
\end{enumerate}
We also say that $S$ (modally) refines $T$ when $S \preceq_m T$.
\end{definition}

Intuitively, $S$ modally refines $T$ if any necessary transition of $T$ can be mimicked by a necessary transition of $S$, and every transition of $S$ can be mimicked by a transition of $T$. 
The set of implementations of an MTS $S$, written $Impl_m(S)$ is defined as the set of LTS $I$ such that $I \preceq_m S$. Indeed, LTS cannot be further refined and are considered implementations. 
In other words,  
every LTS refinement of an MTS $S$ is an
\emph{implementation} of $S$.

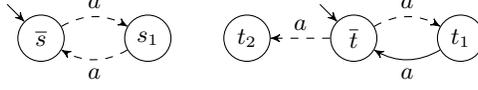
\begin{figure}[tb]
\centering
  \begin{tikzpicture}[>=stealth', every state/.style={draw, minimum 
      size=17.5pt, inner sep=1.5pt}, scale=0.90, node distance=22.5pt] 
    \node[state] (s0) {$\overline s$};
    \node[state, right=of s0] (s1) {$s_1$};
    \draw[->] 
    (s0)++(-0.5,0.5) -- (s0)
    (s0) edge[dashed, out=30, in=150] node[above] {$a$} (s1)   
    (s1) edge[dashed, in=-30, out=-150] node[below] {$a$} (s0); 

    \node[state, right=60pt of s1] (t0) {$\overline t$};
    \node[state, right=of t0] (t1) {$t_1$};
    \node[state, left=of t0] (t2) {$t_2$};
    \draw[->] 
    (t0)++(-0.5,0.5) -- (t0)
    (t0) edge[dashed, out=30, in=150] node[above] {$a$} (t1)   
    (t1) edge[in=-30, out=-150] node[below] {$a$} (t0) 
    (t0) edge[dashed] node[above] {$a$} (t2);

  \end{tikzpicture} 
\caption{\label{fig:mtsrefinement} From left to right, two MTS $S$ and $T$ such that $S \not\preceq_m T$ and $Impl_m(S) \subseteq Impl_m(T)$, showing that modal refinement is not complete (reproduced from~\cite{BKLS12}). Dashed arcs are used to depict optional transitions ($\Delta_\Opt$), while solid arcs depict necessary transitions ($\Delta_{\Box}$). }
\end{figure}

In~\cite{LNW07c}, it is shown that $S \preceq_m T$ implies $Impl_m(S) \subseteq Impl_m(T)$. 
In other words, modal refinement is \emph{sound}, i.e., each time an MTS~$S$ modally refines an MTS~$T$, it follows that the set of implementations of $T$ also contains the implementations of~$S$.  
However, the converse is not true, i.e., modal refinement is not \emph{complete}.
Figure~\ref{fig:mtsrefinement}, reproduced from~\cite{BKLS12}, shows an example where the set of implementations of~$T$ also contains the implementations of~$S$, but~$S$ does not modally refine~$T$.

\section{Non-Reducible MTS Refinement}\label{sect:refinement}

\begin{figure}[tb]
\centering
  \begin{tikzpicture}[>=stealth', every state/.style={draw,  minimum 
      size=17.5pt, inner sep=1.5pt}, scale=0.90, node distance=25pt] 
    \node[state] (s0) {$\overline{s}$};
    \node[state, above right=of s0, xshift=8pt, yshift=-16pt] (s1) {$s_1$};
    \node[state, below right=of s0, xshift=8pt, yshift=16pt] (s2) {$s_2$};
    \node[state, right=of s1] (s3) {$s_3$};
    \node[state, right=of s2] (s4) {$s_4$};
    \node[state, right=of s3] (s5) {$s_5$};
    
    \draw[->] 
    (s0)++(-0.5,0.5) -- (s0)
    (s0) edge[] node[above] {$c$} (s1)   
    (s0) edge[] node[above] {$c$} (s2)
    (s1) edge[dashed] node[above] {$a$} (s3)
    (s2) edge[] node[above] {$a$} (s4)
    (s3) edge[dashed] node[above] {$b$} (s5);

    \node[state, right=155pt of s0] (t0) {$\overline{t}$};
    \node[state, above right=of t0, xshift=8pt, yshift=-16pt] (t1) {$t_1$};
    \node[state, below right=of t0, xshift=8pt, yshift=16pt] (t2) {$t_2$};
    \node[state, right=of t1] (t3) {$t_3$};
    \node[state, right=of t2] (t4) {$t_4$};
    \node[state, right=of t4] (t5) {$t_5$};
    
    \draw[->] 
    (t0)++(-0.5,0.5) -- (t0)
    (t0) edge[] node[above] {$c$} (t1)   
    (t0) edge[] node[above] {$c$} (t2)
    (t1) edge[dashed] node[above] {$a$} (t3)
    (t2) edge[] node[above] {$a$} (t4)
    (t4) edge[dashed] node[above] {$b$} (t5);

  \end{tikzpicture} 
\caption{\label{fig:needCoherence} From left to right, two MTS $S$ and $T$ such that $Impl_m(S) \subseteq Impl_m(T)$ but $S \not\preceq_m T$. 
Both $S$ and $T$ are not NMTS. 
}
\end{figure}
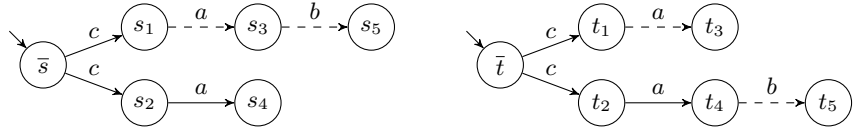

All examples documented in the literature (e.g.,~\cite{BKLS12,LNW07c}), which demonstrate  that modal refinement is not complete (e.g., Figure~\ref{fig:mtsrefinement}), involve the utilization of a non-deterministic choice within the system under refinement. This non-deterministic choice, such as the outgoing transitions from state $\overline t$ in $T$ (as depicted in Figure~\ref{fig:mtsrefinement}), is subsequently eliminated in the refined system, as it is the case in Figure~\ref{fig:mtsrefinement} for system $S$.

Consider Figure~\ref{fig:needCoherence}. 
Similarly to Figure~\ref{fig:mtsrefinement}, it shows an example of two MTS $S$ and $T$ such  that the implementations of $S$ are included into the implementations of $T$, although $S$ and $T$ are not in modal refinement relation. 
Figure~\ref{fig:needCoherence} differs from Figure~\ref{fig:mtsrefinement} (and all other similar examples in the literature) in that it preserves the non-deterministic choices of $T$ within $S$.
In $T$, the states $t_1$ and $t_2$ are reachable by executing the same action $c$ and both exhibit outgoing transitions labeled as $a$. Nonetheless, these two transitions do not share the same modality.

\paragraph{NMTS}
In this section, we identify the subset of MTS that exclusively discards  systems as those in Figure~\ref{fig:needCoherence}.
Indeed, Figure~\ref{fig:needCoherence} shows how such MTS instances can result in a violation of completeness.
We introduce \emph{Non-reducible Modal Transition Systems} (NMTS). 
In NMTS, whenever a sequence of actions leads non-deterministically to different states, all these states are interconnected by the requirement that transitions associated with the same action must also have the same modality. 

In the following, let $w=a_1 \ldots a_n$ be a sequence of actions in $A^*$. The sequence of transitions $\overline s \xrightarrow{a_1} s_1$, $s_1 \xrightarrow{a_2} s_2$, $\ldots$, $s_{n-1} \xrightarrow{a_n} s$ is written as $\overline{s} \xrightarrow{w} s$. 
Furthermore, we write $\overline{s} \not\xrightarrow{w} s$ when it is not possible to reach $s$ from $\overline s$ through the sequence of actions~$w$. 
We say that $w$ is a sequence of $S$ whenever  $\overline{s} \xrightarrow{w} s$. 
To avoid cluttering we denote with $P_S(s)=\{w \mid \overline s \xrightarrow{w} s \}$ the set of sequence of actions of $S$ reaching a state $s \in Q_S$ from the initial state $\overline s$, i.e.,  $\overline{s} \xrightarrow{w} s$ iff $w \in P_S(s)$ and  $\overline{s} \not\xrightarrow{w} s$  iff $w \not\in P_S(s)$. 
We also write $s \xrightarrow{a}$ when the target state of the transition is immaterial.


\begin{definition}[NMTS]\label{def:NMTS}
A \emph{Non-reducible Modal Transition System (NMTS)} $S$ is an MTS  $(Q, A, \overline{s}, \Delta_\Opt, \Delta_\Box)$ such that for all $s, s' \in Q$ whenever $w \in P_S(s) \cap P_S(s')$ there is no $a \in A_S$ such that $s \xrightarrow{a}_\Box$ and $s' \xrightarrow{a}_\Opt$. 
\end{definition}


Clearly, it holds that NMTS are a strict subset of MTS because in NMTS. 


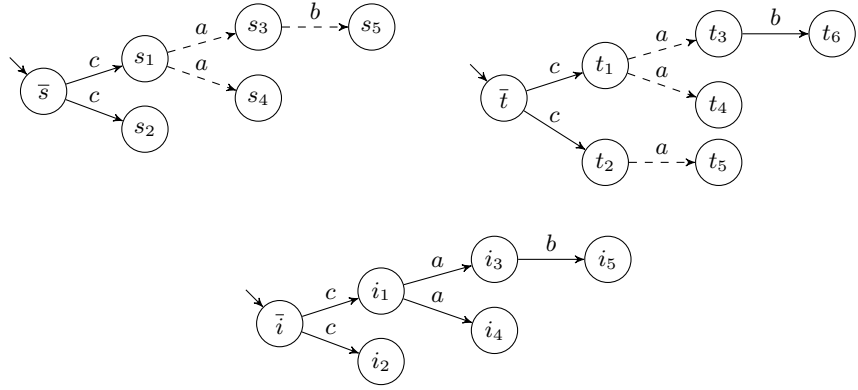
\begin{figure}[tb]
\centering
  \begin{tikzpicture}[>=stealth', every state/.style={draw,  minimum 
      size=17.5pt, inner sep=1.5pt}, scale=0.9, node distance=25pt] 
    \node[state] (s0) {$\overline{s}$};
    \node[state, above right=of s0, xshift=8pt, yshift=-18pt] (s1) {$s_1$};
    \node[state, below right=of s0, xshift=8pt, yshift=16pt] (s2) {$s_2$};
    \node[state, right=of s1,yshift=12pt] (s3) {$s_3$};
    \node[state, below=of s3,yshift=16pt] (s4) {$s_4$};
    \node[state, right=of s3] (s5) {$s_5$};
    
    \draw[->] 
    (s0)++(-0.5,0.5) -- (s0)
    (s0) edge[] node[above] {$c$} (s1)   
        (s0) edge[] node[above] {$c$} (s2)
    (s1) edge[dashed] node[above] {$a$} (s3)
    (s1) edge[dashed] node[above] {$a$} (s4)
    (s3) edge[dashed] node[above] {$b$} (s5);

    \node[state, right=75pt of s4] (t0) {$\overline{t}$};
    \node[state, above right=of t0, xshift=8pt, yshift=-18pt] (t1) {$t_1$};
    \node[state, below right=of t0, xshift=8pt, yshift=6pt] (t2) {$t_2$};
    \node[state, right=of t1, yshift=12pt] (t3) {$t_3$};
    \node[state, below=of t3, yshift=16pt] (t4) {$t_4$};
    \node[state, right=of t2] (t5) {$t_5$};
    \node[state, right=of t3] (t6) {$t_6$};
    
    \draw[->] 
    (t0)++(-0.5,0.5) -- (t0)
    (t0) edge[] node[above] {$c$} (t1)   
    (t0) edge[] node[above] {$c$} (t2)
    (t1) edge[dashed] node[above] {$a$} (t3)
    (t1) edge[dashed] node[above] {$a$} (t4)
    (t2) edge[dashed] node[above] {$a$} (t5)
    (t3) edge[] node[above] {$b$} (t6);
    \end{tikzpicture}
    \phantom{.}\\
    \vspace{0.5cm}
  \begin{tikzpicture}[>=stealth', every state/.style={draw,  minimum 
      size=17.5pt, inner sep=1.5pt}, scale=0.9, node distance=25pt] 
    
    \node[state] (i0) {$\overline{i}$};
    \node[state, above right=of s0, xshift=8pt, yshift=-18pt] (i1) {$i_1$};
    \node[state, below right=of s0, xshift=8pt, yshift=16pt] (i2) {$i_2$};
    \node[state, right=of s1,yshift=12pt] (i3) {$i_3$};
    \node[state, below=of s3,yshift=16pt] (i4) {$i_4$};
    \node[state, right=of s3] (i5) {$i_5$};
    
    \draw[->] 
    (i0)++(-0.5,0.5) -- (i0)
    (i0) edge[] node[above] {$c$} (i1)   
    (i0) edge[] node[above] {$c$} (i2)
    (i1) edge[] node[above] {$a$} (i3)
    (i1) edge[] node[above] {$a$} (i4)
    (i3) edge[] node[above] {$b$} (i5);

  \end{tikzpicture} 
\caption{\label{fig:needConsistence} Top row, from left two right, two NMTS $S$ and $T$ such that  $Impl_m(S) \subseteq Impl_m(T)$ but $S \not\preceq_m T$.  
Furthermore, $S \not\preceq_n T$ and $Impl_n(S) \not\subseteq Impl_n(T)$. Bottom row, an implementation $I$ such that $I \preceq_n S$, $I \not\preceq_n T$, but $I \preceq_m T$}
\end{figure}

Consider Figure~\ref{fig:needConsistence}. In contrast to Figure~\ref{fig:needCoherence}, Figure~\ref{fig:needConsistence} presents two systems, denoted as $S$ and $T$,  satisfying the conditions of Definition~\ref{def:NMTS} (i.e., $S$ and $T$ are NMTS) and satisfying the conditions $Impl_m(S) \subseteq Impl_m(T)$ and $S \not\preceq_m T$. 
Similarly to Figure~\ref{fig:needCoherence}, also 
in Figure~\ref{fig:needConsistence} the non-deterministic choice is maintained. Indeed, in Figure~\ref{fig:needConsistence}, the non-deterministic choice in state $t_1$ in $T$ is maintained in state $s_1$ in $S$. 
Figure~\ref{fig:needConsistence} shows that, for achieving completeness, it is not sufficient to constrain MTS to be NMTS. In the following, we will show that it is also important to introduce  constraints on the refinement relation between NMTS. 

\paragraph{NMTS refinement} 
We now introduce NMTS modal refinement $\preceq_n$. 
In contrast to standard modal refinement, an additional condition is introduced, which applies to the optional transitions within the system undergoing refinement. 
If an optional transition is deactivated during the refinement process, it is required that this deactivation applies uniformly to all other optional transitions sharing the same action. This uniform deactivation rule applies across all source states reachable through the same sequence of actions.

\begin{definition}[NMTS  refinement] \label{def:refinement}
An NMTS  $S$ is an NMTS refinement of another NMTS $T$, denoted as $S \preceq_n T$, if there exists a refinement relation ${\cal R} \subseteq Q_S \times Q_T$ between the states of the two systems such that 
$(\overline s, \overline t) \in \mathcal R$ and for all $(s,t) \in \mathcal R$  it holds that
\begin{enumerate}
    \item 
{whenever} $t \xrightarrow{a}_{\Box} t'$ (for some $t' \in Q_T$, $a \in A_T$), {then} $a \in A_S$ and there exists a state $s' \in Q_S$ such that 
$s \xrightarrow{a}_\Box s'$ and $(s',t') \in {\cal R}$.
\item 
{whenever} $t \xrightarrow{a}_{\Opt} t'$ (for some $t' \in Q_T$, $a \in A_T$), {then} one of the following holds:
\begin{itemize}
\item  $P_S(s) \cap P_T(t) = \emptyset$ or
\item  $P_S(s)  \cap P_T(t) \neq \emptyset$, $\forall s' \in Q_S$\ $\nexists w  \in P_S(s) \cap P_S(s')$ such that  $s' \xrightarrow{a}$; or 
\item $P_S(s) \cap P_T(t) \neq \emptyset$, $a \in A_S$ and there exists a state $s' \in Q_S$ such that $s \xrightarrow{a} s'$ and $(s',t') \in {\cal R}$.
\end{itemize}
\item {whenever} $s \xrightarrow{a} s'$ (for some $s' \in Q_S$, $a \in  A_S$), {then} $a \in A_T$ and there exists a state $t' \in Q_T$ such that $t \xrightarrow{a} t'$ and $(s',t') \in {\cal R}$.
\end{enumerate}
\end{definition}

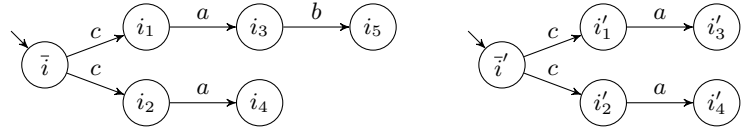
\begin{figure}[tb]
\centering
  \begin{tikzpicture}[>=stealth', every state/.style={draw,  minimum 
      size=17.5pt, inner sep=1.5pt}, scale=0.90, node distance=25pt] 
    \node[state] (0) {$\overline{i}$};
    \node[state, above right=of s0, xshift=8pt, yshift=-16pt] (s1) {$i_1$};
    \node[state, below right=of s0, xshift=8pt, yshift=16pt] (s2) {$i_2$};
    \node[state, right=of s1] (s3) {$i_3$};
    \node[state, right=of s2] (s4) {$i_4$};
    \node[state, right=of s3] (s5) {$i_5$};
    
    \draw[->] 
    (s0)++(-0.5,0.5) -- (s0)
    (s0) edge[] node[above] {$c$} (s1)   
    (s0) edge[] node[above] {$c$} (s2)
    (s1) edge[] node[above] {$a$} (s3)
    (s2) edge[] node[above] {$a$} (s4)
    (s3) edge[] node[above] {$b$} (s5);

    \node[state, right=155pt of s0] (t0) {$\overline{i}'$};
    \node[state, above right=of t0, xshift=8pt, yshift=-16pt] (t1) {$i_1'$};
    \node[state, below right=of t0, xshift=8pt, yshift=16pt] (t2) {$i_2'$};
    \node[state, right=of t1] (t3) {$i_3'$};
    \node[state, right=of t2] (t4) {$i_4'$};
    
    \draw[->] 
    (t0)++(-0.5,0.5) -- (t0)
    (t0) edge[] node[above] {$c$} (t1)   
    (t0) edge[] node[above] {$c$} (t2)
    (t1) edge[] node[above] {$a$} (t3)
    (t2) edge[] node[above] {$a$} (t4);

  \end{tikzpicture} 
\caption{\label{fig:ConsistentNotSufficient} Two LTS $I$ and $I'$ both implementations of the MTS $S$ and $T$ of Figure~\ref{fig:needCoherence}. The set $Impl_n(S)$ contains all and only LTS that are strongly bisimilar to either $I$ or $I'$. 
It follows that $Impl_n(S) \subseteq Impl_n(T)$, and $S \not\preceq_n T$ (under the assumption that $\preceq_n$ is also applicable to MTS)}
\end{figure}

\begin{figure}[tb]
\centering
  \begin{tikzpicture}[>=stealth', every state/.style={draw,  minimum 
      size=17.5pt, inner sep=1.5pt}, scale=0.90, node distance=25pt] 
    \node[state] (0) {$\overline{s}$};
    \node[state, above right=of s0, xshift=8pt, yshift=-16pt] (s1) {$s_1$};
    \node[state, below right=of s0, xshift=8pt, yshift=16pt] (s2) {$s_2$};
    \node[state, right=of s1] (s3) {$s_3$};
    \node[state, right=of s2] (s4) {$s_4$};
    
    \draw[->] 
    (s0)++(-0.5,0.5) -- (s0)
    (s0) edge[] node[above] {$c$} (s1)   
    (s0) edge[] node[above] {$c$} (s2)
    (s1) edge[dashed] node[above] {$a$} (s3)
    (s2) edge[] node[above] {$a$} (s4);

    \node[state, right=155pt of s0] (t0) {$\overline{t}$};
    \node[state, above right=of t0, xshift=8pt, yshift=-16pt] (t1) {$t_1$};
    \node[state, below right=of t0, xshift=8pt, yshift=16pt] (t2) {$t_2$};
    \node[state, right=of t1] (t3) {$t_3$};
    \node[state, right=of t2] (t4) {$t_4$};
    
    \draw[->] 
    (t0)++(-0.5,0.5) -- (t0)
    (t0) edge[] node[above] {$c$} (t1)   
    (t0) edge[] node[above] {$c$} (t2)
    (t1) edge[] node[above] {$a$} (t3)
    (t2) edge[] node[above] {$a$} (t4);

  \end{tikzpicture} 
\caption{\label{fig:NonCoherentFinal} From left to right, two MTS $S$ and $T$ such that $Impl_n(S) \subseteq Impl_n(T)$ and $S \not\preceq_n T$ (under the assumption that $\preceq_n$ is also applicable to MTS). Furthermore,  $Impl_m(S) \not\subseteq Impl_m(T)$ and $S \not\preceq_m T$}
\end{figure}
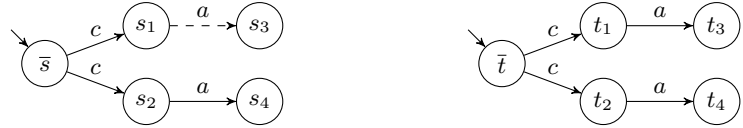


As discussed earlier, Figure~\ref{fig:needConsistence} shows that the further constraint imposed by Definition~\ref{def:NMTS} is not sufficient to achieve completeness of modal refinement. 
Figure~\ref{fig:needCoherence} and Figure~\ref{fig:ConsistentNotSufficient} show that the additional constraint imposed by Definition~\ref{def:refinement} on the refinement relation, when considered independently, is also not sufficient to achieve completeness. 
Indeed, if we switch $\preceq_m$ with $\preceq_n$  in Figure~\ref{fig:needCoherence}, as showed in Figure~\ref{fig:ConsistentNotSufficient}, it would still hold that  $Impl_n(S) \subseteq Impl_n(T)$ and $S \not\preceq_n T$, because $S$ and $T$ are not NMTS.
In other words, the example in 
Figure~\ref{fig:needCoherence} indicates that if non-deterministic MTS do not meet the criteria to be classified as NMTS, then it is possible to build an example, as the one in Figure~\ref{fig:needCoherence}, showing that both modal refinement and NMTS refinement are not complete.

In summary, the examples in Figure~\ref{fig:needCoherence} and Figure~\ref{fig:needConsistence} show that the constraints on MTS and their refinement provided by Definition~\ref{def:NMTS} 
and Definition~\ref{def:refinement} are both needed for achieving completeness. 
By either dropping the constraints on MTS (i.e., Definition~\ref{def:NMTS}) or on their refinement (i.e., Definition~\ref{def:refinement}), it is possible to demonstrate that the resulting refinement relation is not complete. 

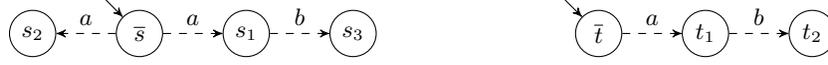
\begin{figure}[t]
\centering
  \begin{tikzpicture}[>=stealth', every state/.style={draw, minimum 
      size=17.5pt, inner sep=1.5pt}, scale=0.90, node distance=22.5pt] 
    \node[state] (s0) {$\overline s$};
    \node[state, right=of s0] (s1) {$s_1$};
    \node[state, left=of s0] (s2) {$s_2$};
    \node[state, right=of s1] (s3) {$s_3$};
    \draw[->] 
    (s0)++(-0.5,0.5) -- (s0)
    (s0) edge[dashed] node[above] {$a$} (s1)   
    (s1) edge[dashed] node[above] {$b$} (s3) 
    (s0) edge[dashed] node[above] {$a$} (s2);

    \node[state, right=75pt of s3] (t0) {$\overline t$};
    \node[state, right=of t0] (t1) {$t_1$};
    \node[state, right=of t1] (t2) {$t_2$};
    \draw[->] 
    (t0)++(-0.5,0.5) -- (t0)
    (t0) edge[dashed] node[above] {$a$} (t1)   
    (t1) edge[dashed] node[above] {$b$} (t2); 
  \end{tikzpicture} 
\caption{\label{fig:example}Two MTS $S$ and $T$ such that $S\preceq_{m} T$, $S \not\preceq_n T$. 
Furthermore,  $T \preceq_m S$ 
 and $T \not\preceq_n S$. In both directions,  
$\overline s \xrightarrow{a}_\Opt s_2$, $\overline t \xrightarrow{a}_\Opt t_1$,  $t_1 \xrightarrow{b}_\Opt t_2$, 
$s_2 \not\xrightarrow{b}$ but $s_1 \xrightarrow{b}$ and both $s_1$ and $s_2$ are reachable by the same sequence of actions
} 
\end{figure}

Figure~\ref{fig:NonCoherentFinal} illustrates another example, proving that the sole constraint imposed by Definition~\ref{def:refinement}  is not sufficient to guarantee completeness. Instead, the constraint required by  Definition~\ref{def:NMTS} is also necessary. 
Note that, differently from Figure~\ref{fig:needCoherence}, Figure~\ref{fig:NonCoherentFinal} is not an example illustrating the incompleteness of $\preceq_m$, because $Impl_m(S) \not\subseteq Impl_m(T)$.

Figure~\ref{fig:example} depicts another example showcasing the differences between $\preceq_m$ and $\preceq_n$.   
Consider the LTS $I_T$ obtained by switching all transitions of $T$ (in Figure~\ref{fig:example}) to must. 
Clearly, $I_T \preceq_n T$, but $I_T \not\preceq_n S$ (note that this is not true for the case of $\preceq_m$). 

\begin{figure}[tb]
\centering
  \begin{tikzpicture}[>=stealth', every state/.style={draw, inner sep=1.5pt}, scale=0.90, node distance=25pt] 
    \node[state] (u0) {$\overline{u}$};
    \node[state, above right=of s0, xshift=8pt, yshift=-16pt] (u) {$u$};
    \node[state, below right=of s0, xshift=8pt, yshift=16pt] (u1) {$u_1$};
    \node[state, right=of u, xshift=-8pt] (u2) {$u_2$};
    \draw[->] 
    (u0)++(-0.5,0.5) -- (u0)
    (u0) edge[] node[above] {$c$} (u1)   
    (u0) edge[bend left=45] node[above] {$c$} (u)
    (u0) edge[] node[above] {$b$} (u)   
    (u) edge[dashed] node[above] {$a$} (u2); 
 
   \node[state, right=90pt of u0] (t0) {$\overline{t}$};
    \node[state, below right=of t0, xshift=8pt, yshift=12pt] (t2) {$t_2$};
    \node[state, right=of t0, xshift=8pt] (t1) {$t_1$};
    \node[state, above right=of t0, xshift=8pt, yshift=-12pt] (t) {$t$};
    \node[state, right=of t1, xshift=-8pt] (t3) {$t_3$};
    \draw[->] 
    (t0)++(-0.5,0.5) -- (t0)
    (t0) edge[] node[above] {$b$} (t)   
    (t0) edge[] node[above] {$c$} (t1)  
    (t0) edge[] node[above] {$c$} (t2)
    (t1) edge[dashed] node[above] {$a$} (t3);

      \node[state, right=110pt of t0] (s0) {$\overline{s}$};
    \node[state, below right=of s0, xshift=8pt, yshift=16pt] (s1) {$s_1$};
    \node[state, above right=of s0, xshift=8pt, yshift=-16pt] (s) {$s$};
    \node[state, right=of s1, xshift=-8pt] (s2) {$s_2$};
    \draw[->] 
    (s0)++(-0.5,0.5) -- (s0)
    (s0) edge[bend left=45] node[above] {$c$} (s)   
    (s0) edge[] node[above] {$b$} (s)
    (s0) edge[] node[above] {$c$} (s1)   
    (s1) edge node[above] {$a$} (s2);

  \end{tikzpicture} 
\caption{\label{fig:notransitivity} From left to right, the NMTS $U$, $T$ and $S$ such that $T \preceq_n U$, $S \preceq_n T$ where 
$\mathcal R_{T \preceq_n U}=\{(\overline t, \overline u), (t_2,u_1),(t,u),(t_1,u),(t_3,u_2)\}$, 
$\mathcal R_{S \preceq_n T}=\{(\overline s, \overline t), (s,t),(s,t_2),(s_1,t_1),(s_2,t_3)\}$.  
When performing $\overline u \xrightarrow{b}_\Box u$, the NMTS $S$ can only reply with 
$\overline s \xrightarrow{b} s$. However, $(s,u)$ violates Definition~\ref{def:refinement}, because 
$u \xrightarrow{a}_\Opt$, $s \not\xrightarrow{a}$ and there exists $c \in P_S(s) \cap P_S(s_1)$ such 
that $s_1 \xrightarrow{a}$. Therefore, $S \not\preceq_n U$. 
Furthermore, since $S$ is an implementation, it holds that $Impl_{n}(T) \not\subseteq Impl_{n}(U)$, 
although $T \preceq_n U$.
}
\end{figure}
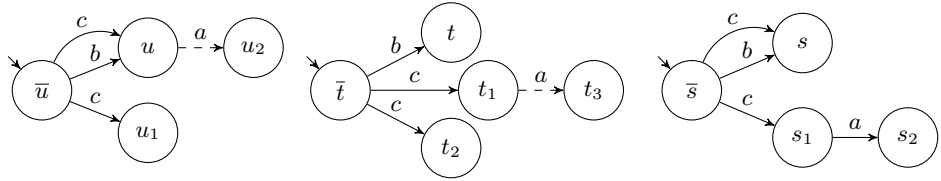

\paragraph{Soundness} The previous examples focus on the problem of achieving completeness and the constraints that might be needed for it. 
However, the current notion of NMTS refinement is not sound (i.e., $S \preceq_n T$ does not imply that $Impl_{n}(S) \subseteq Impl_n(T)$). For example, see Figure~\ref{fig:notransitivity}. Furthermore, currently, there is no formal proof that NMTS refinement in this form is complete.
Therefore, further updates are necessary to either the notion of NMTS or NMTS refinement or both, with the goal of achieving a notion of refinement that is both sound and complete whilst preserving the ability to express non-determinism.

\section{Non-determinism of NMTS is Non-reducible}\label{sect:example}

In the previous section, we discussed the further constraints imposed on MTS and their refinement (i.e., Definition~\ref{def:NMTS} and Definition~\ref{def:refinement}). 
 An important challenge discussed in~\cite{LNW07c} is to argue that the considered set of implementations is also interesting from a practical point of view (i.e., no valuable implementation is disregarded by $\preceq_n$). 
 
In this section we discuss a property concerning non-deterministic optional transitions that is violated by all implementations accepted by $\preceq_m$ and discarded by $\preceq_n$. 
MTS allow to express transitions that must be enabled in all implementations. 
In this case, the presence of non-determinism is unaltered in all implementations, because must transitions cannot be disabled. 
Conversely, optional transitions can be arbitrarily disabled, and in MTS the non-determinism in optional branches can be reduced or fully resolved. 
However, the standard semantics of MTS do not provide the means to specify actions that are susceptible to both enablement and disablement, yet inevitably yield non-deterministic outcomes. 
This issue arises since any action capable of being deactivated (i.e., an optional action) also opens the possibility of diminishing its associated non-determinism.  
We term the property stating that all optional actions in an MTS can be enabled or disabled, while retaining their irreducible non-determinism, as \emph{non-reducible  non-determinism}.

In formal specifications expressed as MTS, non-determinism  is commonly used to express under-specifications. 
This variant of non-determinism does not necessitate preservation across all implementations of an MTS. 
Consequently, modal refinement can reduce the non-determinism to fully determine a specification, i.e., modal refinement 
does not satisfy the non-reducible non-determinism property (see, e.g.,  Figure~\ref{fig:example}). 
There exists a distinction between non-determinism present in all implementations (as showed in the next example) and the non-determinism that characterizes under-specifications. However, both these forms of non-determinism are expressed in an identical way within MTS. 
This inherent ambiguity contributes to the incompleteness of modal refinement.
To address this, we assume that non-deterministic behaviour of MTS is always preserved across all implementations, thereby eliminating non-determinism as a source of under-specification. 
Consequently, NMTS refinement satisfies the property of non-reducible non-determinism, whilst this is not the case for modal refinement.

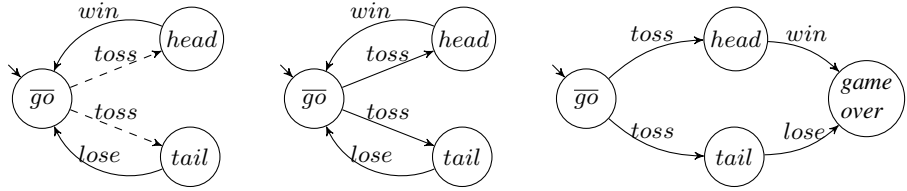
\begin{figure}[tb]
\centering
  \begin{tikzpicture}[>=stealth', every state/.style={draw, inner sep=1.5pt}, scale=0.90, node distance=45pt] 
    \node[state] (s0) {$\overline{go}$};
    \node[state, above right=of s0, xshift=8pt, yshift=-26pt] (s1) {$head$};
    \node[state, below right=of s0, xshift=8pt, yshift=26pt] (s2) {$tail$};
    \draw[->] 
    (s0)++(-0.5,0.5) -- (s0)
    (s0) edge[dashed] node[above] {$toss$} (s1)   
    (s0) edge[dashed] node[above] {$toss$} (s2)   
    (s1) edge[bend right=45] node[above] {$win$} (s0)   
    (s2) edge[bend left=45] node[above] {$lose$} (s0); 
 
    \node[state, right=80pt of s0] (i0) {$\overline{go}$};
    \node[state, above right=of i0, xshift=8pt, yshift=-26pt] (i1) {$head$};
    \node[state, below right=of i0, xshift=8pt, yshift=26pt] (i2) {$tail$};
    \draw[->] 
    (i0)++(-0.5,0.5) -- (i0)
    (i0) edge[] node[above] {$toss$} (i1)   
    (i0) edge[] node[above] {$toss$} (i2)    
    (i1) edge[bend right=45] node[above] {$win$} (i0)   
    (i2) edge[bend left=45] node[above] {$lose$} (i0); 

    \node[state, right=80pt of i0] (j0) {$\overline{go}$};
    \node[state, above right=of j0, xshift=8pt, yshift=-26pt] (j1) {$head$};
    \node[state, below right=of j0, xshift=8pt, yshift=26pt] (j2) {$tail$};
    \node[state, right=80pt of j0, align=left] (j3) {{\it game}\\{\it over}};
    \draw[->] 
    (j0)++(-0.5,0.5) -- (j0)
    (j0) edge[bend left=20] node[above] {$toss$} (j1)   
    (j0) edge[bend right=20] node[above] {$toss$} (j2)    
    (j1) edge[bend left=20] node[above] {$win$} (j3)   
    (j2) edge[bend right=20] node[above] {$lose$} (j3);

  \end{tikzpicture} 
\caption{\label{fig:cointoss} From left to right, the NMTS modelling a coin toss game, an NMTS implementation allowing infinite plays, an NMTS implementation allowing one play. }
\end{figure}

In the following, we discuss an example showcasing the need to establish the non-reducible non-determism property.
Consider the NMTS in Figure~\ref{fig:cointoss} (left). 
This NMTS serves as a model for a coin toss game. 
We visualize the actions of this NMTS as buttons that light up and can only be pressed when the respective action becomes enabled. Upon pressing an enabled button, the associated action is carried out.

Initially, only one action, namely $toss$, is enabled. Upon executing the $toss$ action, the outcome can result in either head or tail. If the outcome is head, the $win$ action becomes enabled, while in the case of tail, the $lose$ action is enabled. Upon the execution of either $win$ or $lose$, the NMTS reverts back to its initial state. 
The NMTS does not specify whether the coin is biased. 

The $toss$  action exemplifies the property of non-reducible non-determinism.
 In essence, any implementation that enables the $toss$ action must consistently manifest the same non-deterministic behavior. Notably, the $toss$ action can be deactivated.
An implementation that restricts the coin's outcomes solely to either heads or tails is considered invalid. However, under the standard modal refinement $\preceq_m$, such invalid implementations are deemed acceptable.
Figure~\ref{fig:cointoss} (center and right) depicts two valid implementations of the NMTS.  In one, an indefinite number of plays are feasible, while in the other, only a single play is permitted.
Both these implementations are preserving the non-deterministic nature  of the $toss$ action.

The introduced NMTS refinement (see Definition~\ref{def:refinement}) exclusively permits implementations like the one showcased in  Figure~\ref{fig:cointoss} whilst forbidding invalid implementations as those forcing the coin to only return either head or tail. 
Clearly, by relaxing the constraints in either Definition~\ref{def:NMTS} or Definition~\ref{def:refinement}, it is possible to define systems whose implementations may violate the non-reducible non-determinism property (see Section~\ref{sect:refinement}).

\section{Related Work}\label{sect:relatedwork}

MTS and their dialects are widely studied in the literature. 
Given two MTS $S$ and $T$, $S$ is a thorough refinement of $T$ whenever the set of implementations of $S$ is included in the set of implementations of $T$. 
In~\cite{LNW07c}, four different refinement relations are studied extensively, including thorough refinement, and an MTS is said to be \emph{consistent} if it admits at least one non-empty implementation. 
MTS that allow inconsistent specifications, where transitions can be necessary but not permitted, are called Mixed Transition Systems~\cite{DGG97,AHLNW08}.   

In~\cite[Corollary~4.6]{BKLS12}, it is proved that, similarly to modal refinement, thorough refinement is decidable in polynomial time for deterministic MTS, whilst thorough refinement is decidable in EXPTIME for non-deterministic MTS. 
The authors describe a tableau-style algorithm~\cite[Section 4]{BKLS12} for deciding thorough refinement, which runs in exponential time 
in the worst case. 
While thorough refinement does not always imply modal refinement of MTS, in~\cite[Lemma~3.6]{BKLS09} it is proved that thorough refinement implies modal refinement of a deterministic overapproximation of (non-deterministic) MTS.

In~\cite[Theorem~3]{LNW07c}, it is proved that any alternative notion $\preceq_{alt}$ of modal refinement that is both sound and complete  cannot be decided in polynomial time unless $\textit{P}\!=\!\textit{NP}$. This is obtained by reducing the problem of deciding thorough refinement to the problem of deciding whether a 3-DNF formula is a tautology. However, in this  case, thorough refinement considers all implementations obtained through modal refinement $\preceq_m$, and not only those obtained using the alternative notion $\preceq_{alt}$. 
The problem of proposing an alternative notion of modal refinement that is both sound and complete with respect to its set of implementations is left open~\cite{LNW07c}. The main challenge is to argue that the considered set of implementations is also interesting from a practical point of view. 
In this paper, we discuss how all implementations retained by $\preceq_m$ and discarded by $\preceq_n$ are violating 
the non-reducible non-determinsm property (see Section~\ref{sect:example}).

Parametric MTS (PMTS)~\cite{BKLMS11,KS13,BKLMSS15} were introduced to enhance the expressiveness of MTS. 
PMTS are LTS equipped with an obligation function~$\Phi$, which is a parametric Boolean proposition over the outgoing transitions from each state. The satisfying assignments of $\Phi$ yield the allowed combinations of outgoing transitions. When~$\Phi$ is not parametric, PMTS are called Boolean MTS (BMTS). PMTS are capable of expressing, among others, \emph{persistent} choices (i.e., once some outgoing transition is enabled, it must be enabled also everywhere else). It is shown that MTS are a special case of BMTS, and that BMTS are a special case of PMTS. 
Rather than extending MTS. 
Thorough refinement is computable in NEXPTIME for both BMTS and PMTS. 
Modal refinement of MTS, BMTS, and PMTS is not complete. 
The deterministic variants of PMTS and BMTS are called, respectively, DPMTS and DBMTS.
When restricting to only deterministic systems, similarly to NMTS, also DBMTS modal refinement is complete, whereas DPMTS modal refinement is still not complete.

In~\cite{BBFG24,BFGM15}, Coherent MTS (CMTS) are introduced as a model for software product lines (SPL). In CMTS, the features of an SPL are identified with the actions of an MTS. 
Therefore, in CMTS  an action cannot be the label of both a necessary and an optional transition, since a feature is either mandatory or optional.  
The notion of \lq consistent\rq\ product derivation requires that whenever an optional transition is discarded in an implementation, all transitions sharing the same label must also be discarded. This consistency requirement mimicks the aforementioned persistency of PMTS~\cite{BKLMS11,KS13,BKLMSS15} and it is not to be confused with the above mentioned notion of consistency as studied in~\cite{LNW07c}. 
In~\cite{BBFG24}  the refinement of CMTS is presented, which is demonstrated to be  both sound and complete in relation to its set of implementations. 

CMTS and their refinement~\cite{BBFG24} are an important milestone in addressing the long-standing problem proposed at CONCUR 2007~\cite{LNW07c}. 
CMTS are currently the only available  subsets of MTS that possess the capacity to preserve both non-deterministic specifications and completeness of the refinement relation. 
In contrast, all the other MTS extensions mentioned above do not   possess completeness of refinement in the case of non-deterministic specifications. 
In CMTS, by interpreting SPL features as MTS actions,  `consistency' and `coherence' are enforced globally across all system states. 
NMTS are a generalization of CMTS. 
In NMTS, the SPL-derived limitations are discarded (i.e., actions are not interpreted as features of an SPL). 
The constraints that in CMTS are applied globally, in NMTS are instead applied exclusively to the set of states reachable through the same sequence of actions. Consequently, NMTS strictly include CMTS. 
Differently from the restrictions imposed by CMTS and their refinement, in Section~\ref{sect:refinement}, we identified the restrictions imposed by NMTS and their refinement as necessary for achieving completeness in the refinement relation. 
Furthermore, in Section~\ref{sect:example} we presented a property that is violated by all implementations discarded by the NMTS refinement relation but accepted through modal refinement.

\section{Conclusion}\label{sect:conclusion}

We have introduced a subset of  Modal Transition Systems (MTS) called  Non-reducible MTS (NMTS) and their refinement relation ($\preceq_n$).

In NMTS,  states reached through the execution of identical action sequences are related. 
Outgoing transitions from related states that are labeled by the same action also exhibit the same modality. 
Disabling an optional transition within a refinement results in the deactivation of all transitions that share both the same action label and are outgoing from related states.
We showed that if either of these conditions is relaxed,  it becomes possible to construct two systems that are not in refinement relation, yet their respective sets of implementations still maintain a relation of set inclusion. 
By interpreting the optional non-determinism present in MTS as non-reducible  (i.e., non-deterministic behaviour within MTS is consistently maintained in all implementations), we have showed that all implementations permitted by $\preceq_m$  (modal refinement) but rejected by~$\preceq_n$  are considered invalid. 

\paragraph{Future work} As reported in Section~\ref{sect:refinement}, the current refinement of NMTS is not sound. Therefore, further research on NMTS and their refinement is needed.


\bibliographystyle{splncs04} 
\bibliography{bib}

\iftoggle{APPENDIX}
{
}
{}

\end{document}